\documentclass[twocolumn,journal]{IEEEtran}
\usepackage[T1]{fontenc}
\usepackage[latin9]{inputenc}
\usepackage{color}
\usepackage{amsmath}
\usepackage{amssymb}
\usepackage{graphicx}
\usepackage[unicode=true,
 bookmarks=true,bookmarksnumbered=true,bookmarksopen=true,bookmarksopenlevel=1,
 breaklinks=false,pdfborder={0 0 0},pdfborderstyle={},backref=false,colorlinks=false]
 {hyperref}
\hypersetup{pdftitle={Your Title},
 pdfauthor={Your Name},
 pdfpagelayout=OneColumn, pdfnewwindow=true, pdfstartview=XYZ, plainpages=false}

\makeatletter
\let\oldforeign@language\foreign@language
\DeclareRobustCommand{\foreign@language}[1]{%
  \lowercase{\oldforeign@language{#1}}}

\usepackage[caption=false,font=footnotesize]{subfig}

\@ifundefined{showcaptionsetup}{}{%
 \PassOptionsToPackage{caption=false}{subfig}}
\usepackage{subfig}
\makeatother

\begin{document}
\title{Adaptive Multi-User Channel Estimation Based on Contrastive Feature
Learning}
\author{Yihan Xu,~\IEEEmembership{Student Member,~IEEE,} Lixiang Lian,~\IEEEmembership{Member,~IEEE}\thanks{This work is sponsored by National Natural Science Foundation of China
(NSFC) under Grant 62101331 \textit{(Corresponding author: Lixiang
Lian)}.}\thanks{Yihan Xu is with the School of Information Science and Technology,
ShanghaiTech University, Shanghai, China, e-mail: \protect\href{mailto:xuyh3@shanghaitech.edu.cn}{xuyh3@shanghaitech.edu.cn}.}\thanks{Lixiang Lian is with the School of Information Science and Technology,
ShanghaiTech University, Shanghai, China, e-mail: \protect\href{mailto:lianlx@shanghaitech.edu.cn}{lianlx@shanghaitech.edu.cn}.}}
\markboth{}{XU \MakeLowercase{\emph{et al.}}: Adaptive Multi-User Channel Estimation
Based on Contrastive Feature Learning}
\maketitle
\begin{abstract}
Correlation exploitation is essential for efficient multi-user channel
estimation (MUCE) in massive MIMO systems. However, the existing works
either rely on presumed strong correlation or learn the correlation
through large amount of labeled data, which are difficult to acquire
in a real system. In this paper, we propose an adaptive MUCE algorithm
based on contrastive feature learning. The contrastive learning (CL)
is used to automatically learn the similarity within channels by extracting
the channel state information (CSI) features based on location information.
The similar features will be fed into the downstream network to explore
the strong correlations among CSI features to improve the MUCE performance
with a small number of labeled data. Simulation results show that
the contrastive feature learning can enhance the overall MUCE performance
with high training efficiency.
\end{abstract}


\IEEEpeerreviewmaketitle{}

\section{Introduction}

\IEEEPARstart{T}{he} fifth generation (5G) and 5G beyond wireless
networks are characterized by extensive use of massive multiple-input
multiple-output (MIMO) techniques \cite{larsson2014massiveMIMO,swindlehurst2014mmwavemassiveMIMO}.
In massive MIMO, channel estimation (CE) plays a vital role for advanced
transmission scheme designs. However, it is challenging to acquire
accurate channel state information (CSI) at base station (BS) in a
frequency-division duplexing (FDD) system due to the large scale of
CSI and the loss of channel reciprocity. Many studies have shown that
the channels of multi-user (MU) massive MIMO systems exhibit certain
location-dependent spacial correlations due to the shared transmission
environment \cite{ko2012multiuser_distance,le2021similarity}. Various
compressive sensing (CS)-based algorithms have been developed to exploit
the common sparsity of channels among nearby users \cite{rao2014distributed,liu2018downlink,Liu2019bayesian}.
However, traditional MUCE algorithms highly depend on some restrictive
assumptions (e.g., channel sparsity, common sparsity, channel priors),
and are iterative and computational intensive, which hinder their
applications in practical systems.

Machine learning (ML) has been leveraged to develop efficient CE algorithms
\cite{soltani2019dlforCE,kang2018dlCEwireless,chun2019dlCEMIMO}.
Most of works focused on supervised approaches, where the quantity
and quality of label data directly affect the final results of ML
algorithms. In CE problem, the labeled data is the ground-truth channel,
which is usually hard to measure or is hardware-intensive to acquire
in practice. To improve the training efficiency and enhance the adaptability
of the neural network (NN) to the actual channel condition, it is
important to develop advanced ML-based CE (MCE) algorithm that can
learn effectively from a small number of labeled data. Moreover, for
MUCE problem, existing works \cite{su2021multiUE_CE,chun2019multiuserCE}
leave the NN to exploit the internal correlations among channels in
a data-driven way by directly feeding the MU data into the NN, which
is highly inefficient and can lead to performance loss when training
data is insufficient.

In this paper, we propose an adaptive MUCE algorithm based on contrastive
learning (CL) \cite{le2020contrastivereview}. Motivated by the fact
that CSI contains location information of user \cite{garcia2017direct_localization}
and two channel matrices measured at adjacent positions should be
close to each other through a specific similarity metric \cite{sobehy2020csi_knn,le2021similarity_position},
we generate positive and negative samples based on users\textquoteright{}
location information to train a CLNet, such that the input data collected
from adjacent locations are mapped by the CLNet to similar CSI features
in an intermediate feature space. Then similar features will be fed
into a downstream network (DNet) to exploit the correlations among
the input features to improve the CE performance. The joint learning
of CLNet and DNet can further refine the features extracted by the
CLNet in a task-oriented way. The proposed framework offers the following
advantages: 1) The CLNet can be trained with only unlabeled data.
Benefiting from the contrastive feature learning, DNet can be well
trained with a small number of labeled data. 2) In the training phase,
CLNet can automatically learn the similarities among MU channels,
which can be exploited by the DNet to improve the CE performance.
In the testing phase, CSI similarities of MU data can be checked using
trained CLNet, based on which, appropriate DNet can be chosen to perform
MUCE.

\section{System Model}

\begin{figure}[htbp]
\begin{centering}
\textsf{\includegraphics[scale=0.6]{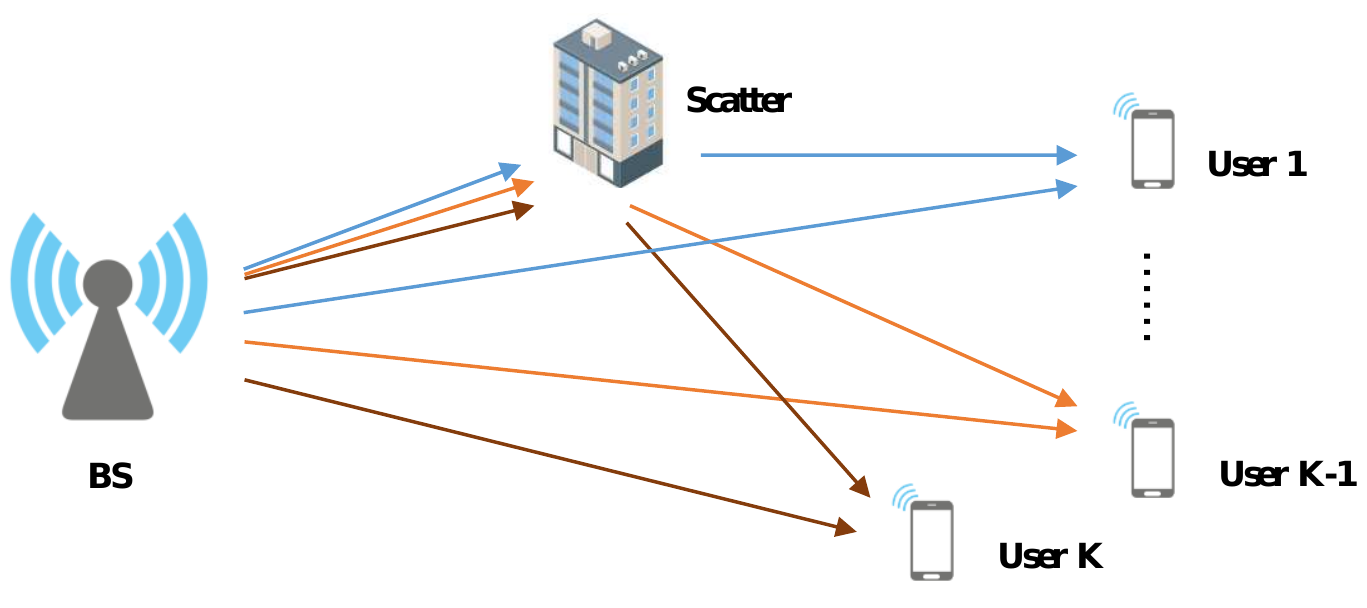}}
\par\end{centering}
\caption{System model of MUCE in massive MIMO systems.\label{fig:System-Model}}
\end{figure}

Consider an orthogonal frequency division multiplexing (OFDM) massive
MIMO system with one BS serving $K$ users, as illustrated in Fig.
\ref{fig:System-Model}. The BS is equipped with $N_{t}$ antennas.
Each user is equipped with $N_{r}$ antennas. There are $N_{c}$ subcarriers.
The BS broadcasts the pilot signal $\mathbf{\mathbf{S}}_{n}\in\mathbb{C}^{N_{t}\times L}$
at the $n$-th subcarrier for downlink CE, where $L$ is the pilot
length. Then the received signal at the $k$-th user and $n$-th subcarrier
is given by

\begin{equation}
\mathbf{Y}_{k,n}=\mathbf{H}_{k,n}\mathbf{S}_{n}+\mathbf{N}_{k,n},
\end{equation}
where $\mathbf{H}_{k,n}\in\mathbb{C}^{N_{r}\times N_{t}}$ is the
channel matrix from BS to the $k$-th user at the $n$-th subcarrier
and $\mathbf{N}_{k,n}$ is the additive complex Gaussian noise (AWGN)
with each element being zero mean and variance $\sigma^{2}$. Denote
$\mathbf{H}_{k}=[\mathbf{H}_{k,1},\cdots,\mathbf{H}_{k,N_{c}}]\in\mathbb{C}^{N_{r}\times N_{t}N_{c}}$
and $\mathbf{Y}_{k}=[\mathbf{Y}_{k,1},\cdots,\mathbf{Y}_{k,N_{c}}]\in\mathbb{C}^{N_{r}\times LN_{c}}.$
The goal of MUCE is to jointly estimate $\{\mathbf{H}_{k}\}$ from
$\{\mathbf{Y}_{k}\}$ exploiting the hidden spatial correlations of
the channel matrices to save the pilot overhead. In the rest of this
paper, \textquoteleft data' means the measurements $\{\mathbf{Y}_{k}\}$
used for contrastive learning, \textquoteleft labeled data' means
the ground-truth channel $\{\mathbf{H}_{k}\}$ used for downstream
network learning.

\section{Contrastive Feature Learning in MUCE}

\begin{figure*}[tp]
\begin{centering}
\textsf{\includegraphics[scale=0.5]{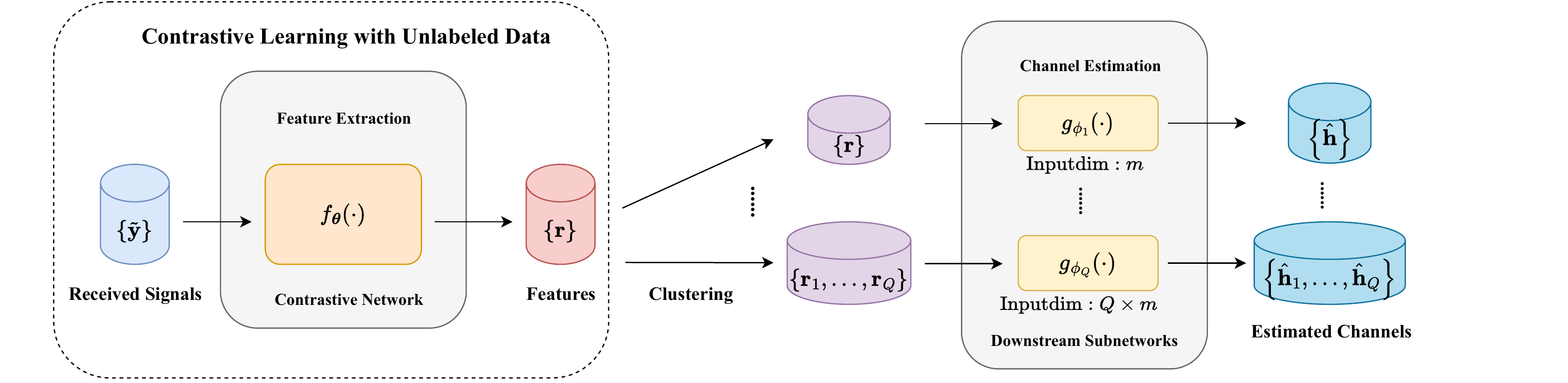}}
\par\end{centering}
\centering{}\caption{The proposed network architecture of cascaded CLNet-DNet, where the
CLNet uses unlabeled data for training to complete feature extraction
and DNet exploits the similar CSI features for joint MUCE with reduced
number of labeled data. \label{fig:Network}}
\end{figure*}

CL has been used to solve resource allocation problems in wireless
communication systems, such as power control \cite{naderializadeh2021contrastive_power_control}
and localization \cite{deng2022supervised}. In this paper, we propose
a CL-based MUCE algorithm. The overall network architecture is illustrated
in Fig. \ref{fig:Network}. By selecting the generation criteria for
positive and negative samples and using the contrastive loss function,
features that are helpful for downstream task can be learned from
unlabeled data through CLNet. After that, the learned features are
used as input to the DNet to explore the hidden correlations among
channels with less labeled data. In this section, we first introduce
a feature extraction network based on CL. Then we introduce the DNet
to establish a mapping between the features and the estimated channels.
At the end, we introduce an adaptive MUCE framework.

\subsection{Contrastive Network}

Assume the mapping function from the received signals to features
is $f_{\boldsymbol{\theta}}(\cdot)$ with model parameters $\boldsymbol{\theta}$.
Considering that the channel matrix is often composed of complex numbers,
we establish a mapping from complex vector to real vector as
\begin{equation}
\nu:\mathbf{z}\in\mathbb{C}^{M}\to\tilde{\mathbf{z}}=\left[\mathrm{Re}(\mathbf{z})^{T},\mathrm{Im(\mathbf{z})}^{T}\right]^{T}\in\mathbb{R}^{2M}.
\end{equation}
Then, the feature extraction of CLNet can be expressed as
\begin{equation}
\mathbf{r}=f_{\boldsymbol{\theta}}(\nu(\mathbf{y}))=f_{\boldsymbol{\theta}}(\tilde{\mathbf{y}})\in\mathbb{R}^{m},
\end{equation}
where $\mathbf{y}=\mathrm{vec}(\mathbf{Y}),$ $\mathbf{r}$ represents
the feature vector obtained from the received signal $\mathbf{y}$
and $m$ is the dimension of feature vector. For brevity, we omit
the user index here.

We design a deep convolutional NN (CNN) to learn the CSI features,
which is trained using CL method. The goal of CL is that the features
of two similar data should end up close to each other in the $m$-dimensional
feature space, while the features for two different data should get
as far as possible from each other. To learn the similarities inherent
in CSI, we regard the received signals from adjacent locations as
similar data. Consider a training dataset $\{\tilde{\mathbf{y}}_{i},\mathbf{p}_{i}\}_{i=1}^{I_{C}}$,
where $\mathbf{p}_{i}$ represents the position of user with received
signal $\tilde{\mathbf{y}}_{i}$. For a received signal sample $\tilde{\mathbf{y}}_{i}$,
its positive samples are constructed as the set of $d$-nearest data
in the training dataset. $d$-nearest data is defined as follows:
if $\tilde{\mathbf{y}}_{i^{*}}$ is a $d$-nearest data of $\tilde{\mathbf{y}}_{i}$,
then their corresponding positions satisfy $\|\mathbf{p}_{i^{*}}-\mathbf{p}_{i}\|_{2}\leq d$.
The negative samples of $\tilde{\mathbf{y}}_{i}$ are randomly selected
from the remaining data, which means the Euclidean distance between
their corresponding positions is larger than $d$. Denote the positive
samples of $\tilde{\mathbf{y}}_{i}$ are $\{\tilde{\mathbf{y}}_{a}\}_{a\in\mathcal{A}_{i}}$
and its negative samples are $\{\tilde{\mathbf{y}}_{b}\}_{b\in\mathcal{B}_{i}}$.
Then the CLNet can be trained to minimize the following contrastive
loss:
\begin{equation}
L_{\mathrm{CL}}=\frac{1}{I}\sum_{i=1}^{I}\left[-\sum_{a\in\mathcal{A}_{i}}\mathrm{log}\frac{\mathrm{s}(\mathbf{r}_{i},\mathbf{r}_{a})}{\sum_{a\in\mathcal{A}_{i}}\mathrm{s}(\mathbf{r}_{i},\mathbf{r}_{a})+\sum_{b\in\mathcal{B}_{i}}\mathrm{s}(\mathbf{r}_{i},\mathbf{r}_{b})}\right],
\end{equation}
where $\mathbf{r}_{i}$, $\mathbf{r}_{a}$ and $\mathbf{r}_{b}$ are
the features of data $\tilde{\mathbf{y}}_{i}$, its positive samples
and negative samples, respectively. $\mathrm{s(\cdot)}$ is used to
measure the similarity between two vectors, given by
\begin{equation}
\mathrm{s}(\mathbf{r}_{i},\mathbf{r}_{i^{'}})=e^{\mathbf{r}_{i}\cdot\mathbf{r}_{i^{'}}/\tau},
\end{equation}
where $\tau$ is a scalar temperature hyperparameter \cite{robinson2020contrastive}.
The inner product-based similarity metric is more stable and efficient
compared to $\ell_{2}$-norm-based metric \cite{he2020momentum}.
Unlike common contrastive loss functions, such as InfoNCE loss \cite{oord2018representation},
the contrastive loss function we used involves multiple positive samples
to extract location-embedded CSI features more efficiently. Note that
although the dimension of the received signal $\mathbf{y}$ decreases
as the pilot length decreases, we can select the appropriate value
of $m$ so that the feature $\mathbf{r}$ contains enough information
for CE.

For a well trained CLNet $f_{\boldsymbol{\theta}}(\tilde{\mathbf{y}})$,
the CSI similarity metric can be constructed as
\begin{equation}
\gamma\left(\tilde{\mathbf{y}}_{i},\tilde{\mathbf{y}}_{j}\right)=\left\Vert f_{\boldsymbol{\theta}}(\tilde{\mathbf{y}}_{i})-f_{\boldsymbol{\theta}}(\tilde{\mathbf{y}}_{j})\right\Vert _{2}^{-1}=\left\Vert \mathbf{r}_{i}-\mathbf{r}_{j}\right\Vert _{2}^{-1},\label{eq:similarity}
\end{equation}
which will be used to cluster the training data samples to train the
downstream network.

\subsection{Downstream Network\label{subsec:Downstream-Network}}

Assume the mapping function from the features to the estimated channels
is $g_{\boldsymbol{\phi}}(\cdot)$ with model parameters $\boldsymbol{\phi}$.
In order to output channel matrix, we construct the following inverse
mapping:
\begin{equation}
\nu^{-1}:\tilde{\mathbf{z}}=\left[\mathrm{Re}(\mathbf{z})^{T},\mathrm{Im(\mathbf{z})}^{T}\right]^{T}\in\mathbb{R}^{2M}\to\mathbf{z}\in\mathbb{C}^{M}.
\end{equation}
Hence, the CE at the DNet can be expressed as
\begin{equation}
\hat{\mathbf{h}}=\nu^{-1}(g_{\boldsymbol{\phi}}(\mathbf{r})).
\end{equation}

CNN-based DNet is utilized to complete downstream task, which is trained
in a supervised learning approach with a small number of labeled data.
For CSI with certain similarities, the inherent spatial correlations
can be exploited to enhance the overall CE performance and reduce
the training overhead. We construct the downstream training dataset
exploiting the similarity metric learned by CLNet to guarantee that
the input features of DNet share high similarities. Assume the total
training dataset of DNet is given by $\{\mathbf{y}_{i},\mathbf{h}_{i}\}_{i=1}^{I_{D}}$,
where $\mathbf{h}_{i}=\mathrm{vec}(\mathbf{H}_{i})$ is the labeled
channel data. We pass the measurement samples $\{\mathbf{y}_{i}\}$
through the well trained CLNet to obtain their CSI features $\{\mathbf{r}_{i}\}$.
Then the feature-channel data samples $\{\mathbf{r}_{i},\mathbf{h}_{i}\}_{i=1}^{I_{D}}$
are further clustered into $T$ groups based on the CSI similarity
metric proposed in (\ref{eq:similarity}), such that each group contains
highly similar features. Assume each group contains $J$ pairs of
data, then the training dataset can be written in matrix forms as
$\{\mathbf{R}_{t},\breve{\mathbf{H}}_{t}\}_{t=1}^{T}$, where $\mathbf{R}_{t}\in\mathbb{R}^{m\times J}$
is composed of $J$ feature vectors in the $t$-th group, and $\breve{\mathbf{H}}_{t}\in\mathbb{C}^{N_{r}N_{t}N_{C}\times J}$
is composed of $J$ channel vectors in the $t$-th group. The objective
of the supervised learning for DNet is to minimize the mean squared
error (MSE):
\begin{equation}
\mathcal{L}_{\mathrm{MSE}}=\frac{1}{T}\sum_{i=1}^{T}\|\mathbf{\breve{\mathbf{H}}}_{t}-\nu^{-1}(g_{\boldsymbol{\phi}}(\mathbf{R}_{t}))\|_{F}^{2}.
\end{equation}

\subsection{Joint Learning}

The CLNet and DNet can be trained separately. They also can be trained
jointly to refine the feature extraction of CLNet in a task-oriented
manner to serve the final CE task. The training dataset can be preprocessed
using pretrained CLNet and divided into $T$ clusters, each with $J$
data pairs. The whole mapping function of cascaded CLNet-DNet can
be expressed as

\begin{equation}
\hat{\mathbf{h}}=\nu^{-1}\left(g_{\boldsymbol{\phi}}\left(f_{\boldsymbol{\theta}}\left(\nu(\mathbf{y})\right)\right)\right).
\end{equation}

The model parameters $\boldsymbol{\theta}$ and $\boldsymbol{\phi}$
can be learned by minimizing the following loss function:

\begin{equation}
\begin{aligned} & \mathcal{L}_{\mathrm{Joint}}=\\
 & \frac{1}{T}\sum_{t=1}^{T}\left(\alpha\cdot\mathcal{L}_{\mathrm{sim}}+\|\mathbf{\breve{H}}_{t}-\nu^{-1}\left(g_{\boldsymbol{\phi}}\left(\left\{ f_{\boldsymbol{\theta}}\left(\tilde{\mathbf{y}}_{t,i}\right)\right\} _{i=1}^{J}\right)\right)\|_{F}^{2}\right)
\end{aligned}
\end{equation}
where $\alpha\in\mathbb{R}_{+}$ is a hyperparameter and $\mathcal{L}_{\mathrm{sim}}$
characterizes the similarity of the data in each cluster, which is
defined as
\begin{equation}
\mathcal{L}_{\mathrm{sim}}=-\sum_{j=1}^{J-1}\sum_{i=j+1}^{J}\mathrm{s}(f_{\boldsymbol{\theta}}(\tilde{\mathbf{y}}_{t,i}),f_{\boldsymbol{\theta}}(\tilde{\mathbf{y}}_{t,j})).
\end{equation}

\subsection{Adaptive MUCE}

To design an adaptive MUCE algorithm which can adapt to different
degrees of similarities, we propose a subnetwork-based framework,
as illustrated in Fig. \ref{fig:Network}. Specifically, we construct
$Q$ downstream subnetworks (DSNets), each of which is trained in
a similar manner as expressed in Subsection \ref{subsec:Downstream-Network},
but with different number of input features. Based on the similarity
metric $\gamma$ in (\ref{eq:similarity}), the total training dataset
$\{\mathbf{y}_{i},\mathbf{h}_{i}\}_{i=1}^{I_{D}}$ of downstream task
can be divided into different sized clusters. The clustered training
dataset with cluster size equal to $q$ is denoted as $\mathcal{D}_{q},$
where $q=1,\cdots,Q$. Then $\mathcal{D}_{q}$ can be used to train
the $q$-th DSNet which has $q$ input features. To improve the overall
performance, the $Q$-th DSNet with the most number of input features
will be jointly trained with the CLNet to refine the feature extraction
process. Then the remaining $Q-1$ DSNets are separately trained based
on the well retrained CLNet. In the test phase, given $K$ users'
received signals $\{\mathbf{y}_{k}\}_{k=1}^{K}$, we firstly feed
them into the well trained CLNet to get the CSI features $\{\mathbf{r}_{k}\}.$
Then CSI similarity can be checked based on similarity metric $\gamma$
in (\ref{eq:similarity}), after which, $K$ users will be divided
into several groups to ensure that the CSI features within each group
shares a relatively strong similarity. According to the size of each
group, the collected features will be fed into the corresponding DSNet
to obtain the final CE results. Through such sub-network design, the
proposed scheme can realize automatic user clustering in a task oriented
manner only based on the measurement signals, such that the users
within each cluster can efficiently assist each other to achieve a
good CE performance. Moreover, the proposed scheme can be applied
to any number of users.

\subsection{Implementations}

\textcolor{blue}{ }In FDD systems, due to the lack of reciprocity
between downlink and uplink CSI, the acquisition of downlink CSI usually
consists of two phases. In phase one, the BS sends the pilot signal
to UE for downlink channel estimation and UE receives the measurements.
In phase two, the UE will feedback the received measurements to the
BS, so that the BS can estimate the downlink CSI from the feedback
measurements. In this paper, we consider that the feedback measurements
from users are collected at the BS, based on which, the proposed MUCE
algorithm is executed at the BS. Therefore, the process of feature
extraction from the measurements, user clustering based on the similarity
score and the feeding of the similar features to the DNet are all
performed at the BS, which has all the required information.

\section{Simulation Results}

We verify the performance of proposed scheme using real outdoor massive
MIMO dataset provided by \cite{gauger2020dataset}. Consider $N_{t}=56$
and single-antenna users. The OFDM channel has a bandwidth of 20 MHz
with 1024 sub-carriers, and we focus on the CE for first subcarrier
without loss of generality. The pilot matrix is set to be i.i.d. normal
distributed. The CLNet is trained with $I_{C}=4979$ received signal
samples. The DNet is trained with $I_{D}=1500$ labeled data.

The CLNet consists of four sets of convolution operations with parameters
$ker=[4,2,2,2]$, $str=[2,1,1,1]$, $pad=[1,1,1,0]$, $cha=[8,16,16,32]$,
where $ker$, $str$, $pad$ and $cha$ correspond to the respective
kernel, stride, padding sizes and the number of channels of feature
maps. Then, two fully connected layers are added to output CSI features,
where the last layer is with dimension $m=2N_{r}N_{t}N_{c}$. DSNets
consist of five sets of convolution operations with parameters $ker=[q,2,2,2,2]$,
$str=[2,1,1,1,1]$, $pad=[1,1,1,0,0]$ and $cha=[8,16,32,64,64]$,
where $q=1,\cdots,Q$. We set $Q=3$ in the training of subnetworks.
In the test phase, we estimate the channels of $K=5$ users jointly
using the proposed CLNet and DSNets framework. The weights of NNs
are updated by Adam optimizer, whose initial learning rate is 0.0001
and weight decay is 0.01.

\begin{figure}[tbh]
\begin{centering}
\includegraphics[scale=0.5]{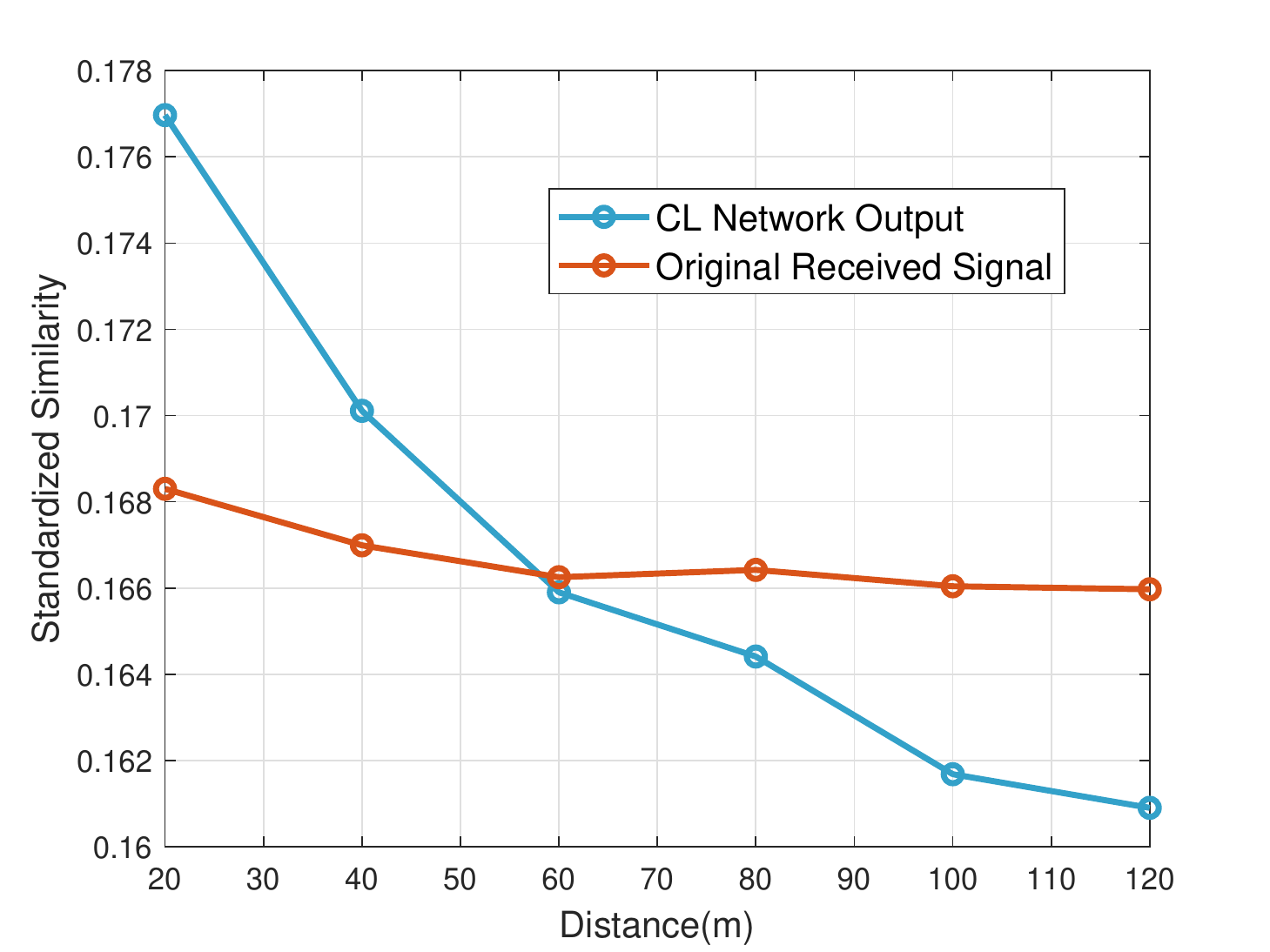}\caption{Standardized similarity among received signals versus distance.\label{fig:Standardized-similarity}}
\par\end{centering}
\end{figure}

First, in order to verify that the output features of CLNet are embedded
with the CSI similarity information, we compare the similarity (defined
as (\ref{eq:similarity})) between original received signals with
that between CSI features for different distances. The curve of original
received signal is obtained by substituting the feature vector $\mathbf{r}_{i}$
in (\ref{eq:similarity}) with the vectorized received signal $\tilde{\mathbf{y}}_{i}$,
that is the similarity between the received signals is calculated
by $\left\Vert \tilde{\mathbf{y}}_{i}-\tilde{\mathbf{y}}_{j}\right\Vert _{2}^{-1}$\textcolor{blue}{.}
As shown in Fig.\ref{fig:Standardized-similarity}, original received
signal does not present explicit location-related information and
cannot reflect the similarity of the channels. Thus, it\textquoteright s
hard to exploit the channel correlations directly from the received
signals. On the contrary, the CSI features bear the location-related
information explicitly and the similarities between CSI features can
effectively reflect the correlations between channels. Therefore,
the proposed CLNet successfully learns the inherent CSI similarities
to assist the downstream MUCE task.

\begin{figure*}[tbh]
\begin{centering}
\textsf{}\subfloat[\label{fig:SNR}]{\begin{centering}
\includegraphics[width=0.32\textwidth]{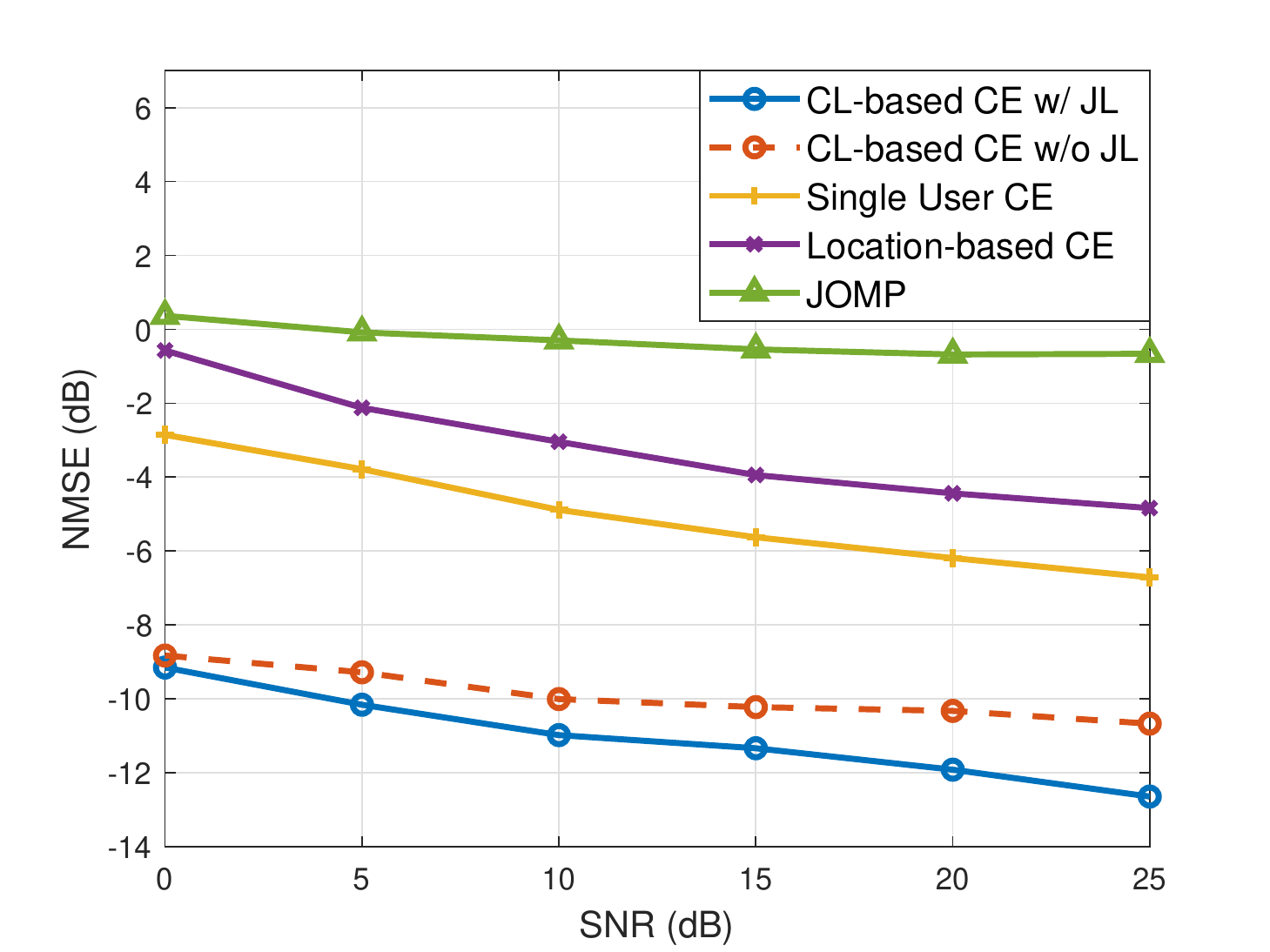}
\par\end{centering}
\raggedright{}\textsf{}}\textsf{}\subfloat[\label{fig:pilot length}]{\begin{centering}
\includegraphics[width=0.32\textwidth]{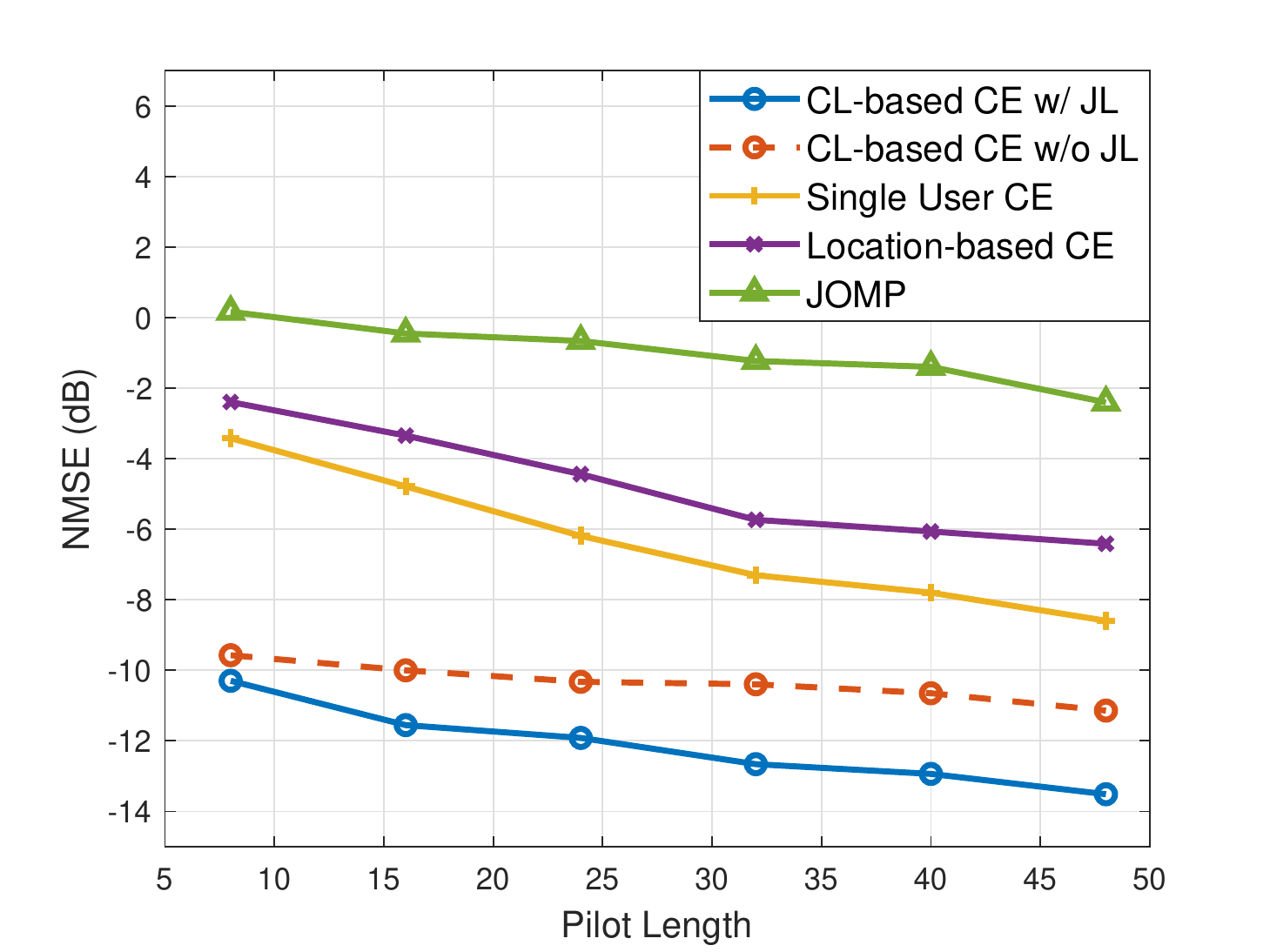}
\par\end{centering}
\begin{raggedright}
\textsf{}
\par\end{raggedright}
}\textsf{}\subfloat[\label{fig:labeled data number}]{\begin{centering}
\includegraphics[width=0.32\textwidth]{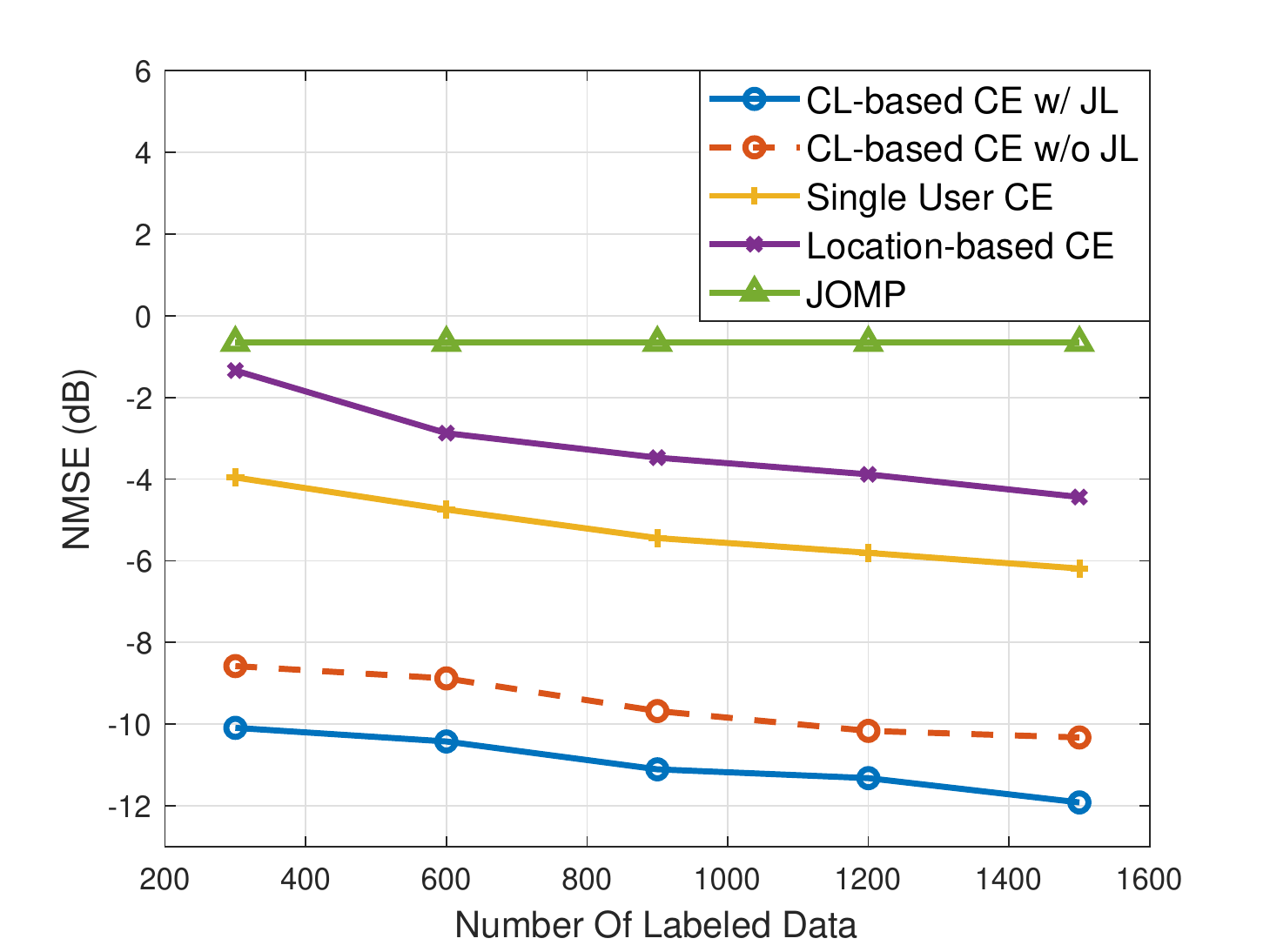}
\par\end{centering}
\begin{raggedright}
\textsf{}
\par\end{raggedright}
}
\par\end{centering}
\caption{(a) NMSE versus SNR with $L=24$ and 1500 labeled data. (b) NMSE versus
pilot length with SNR = 20dB and 1500 labeled data. (c) NMSE versus
number of labeled data with $L=24$ and SNR = 20dB.}
\end{figure*}

We use normalized mean square error (NMSE) defined as $\frac{1}{N}\sum_{n=1}^{N}\|\mathbf{h}_{n}-\tilde{\mathbf{h}}_{n}\|_{2}^{2}/\|\mathbf{h}_{n}\|_{2}^{2}$
to evaluate the CE performance. Consider the following baselines:
1) Single User CE, which trains the CNN in a supervised way to learn
the channel for each user from the received signal individually; 2)
Location-based CE, where the users are firstly clustered based on
users' locations using traditional clustering algorithms, such as
K-Means, then a CNN is trained in a supervised way to jointly learn
the channels of users in a cluster from the received signals; 3) A
Joint Orthogonal Matching Pursuit (JOMP) method proposed in \cite{rao2014distributed},
which exploits the hidden joint sparsity in the user channel matrices
to complete MUCE. We verify the performances of the proposed CL-based
CE with and without the joint CLNet-DNet learning. For joint learning,
the hyperparameter $\alpha$ of $\mathcal{L}_{\mathrm{Joint}}$ is
equal to 0.8.

In Fig. \ref{fig:SNR}, we test the performances of different algorithms
under different SNR values. It shows that the proposed scheme can
achieve the best CE performance and joint learning can further boost
the overall performance. The location-based CE performs the worst,
which can be explained as follows. Due to the presence of noise as
well as the limited pilot sequences, i.e., $L<N_{t}$, the received
signals do not bear the CSI correlation information explicitly and
therefore the differences between them can be large even if they are
close in position. Consequently, this baseline scheme cannot effectively
exploit the correlations among MU channels especially with scarcity
of labeled data, which leads to performance loss. Moreover, when the
similarities between input signals are relatively low, the joint MUCE
can negatively affect each other, resulting in worse performance than
SUCE.

In Fig. \ref{fig:pilot length}, we test the performance of the proposed
algorithm for different pilot lengths. It shows that the proposed
scheme can achieve the best CE performance with very small number
of pilot sequences. Therefore, the CLNet can significantly reduce
the pilot overhead for CE and improve the communication efficiency
of the whole system.

In Fig. \ref{fig:labeled data number}, we verify the training efficiency
of different algorithms. It shows that our proposed CL-based CE algorithm
can significantly reduce the number of labeled data for downstream
network training, thereby achieving high training efficiency. The
high efficiency also enables the deployment of proposed scheme on
a practical system to adapt to the real transmission environment more
efficiently.

\begin{figure}[tbh]
\subfloat[Locations of measurements.\label{fig:Locations}]{\begin{centering}
\includegraphics[scale=0.24]{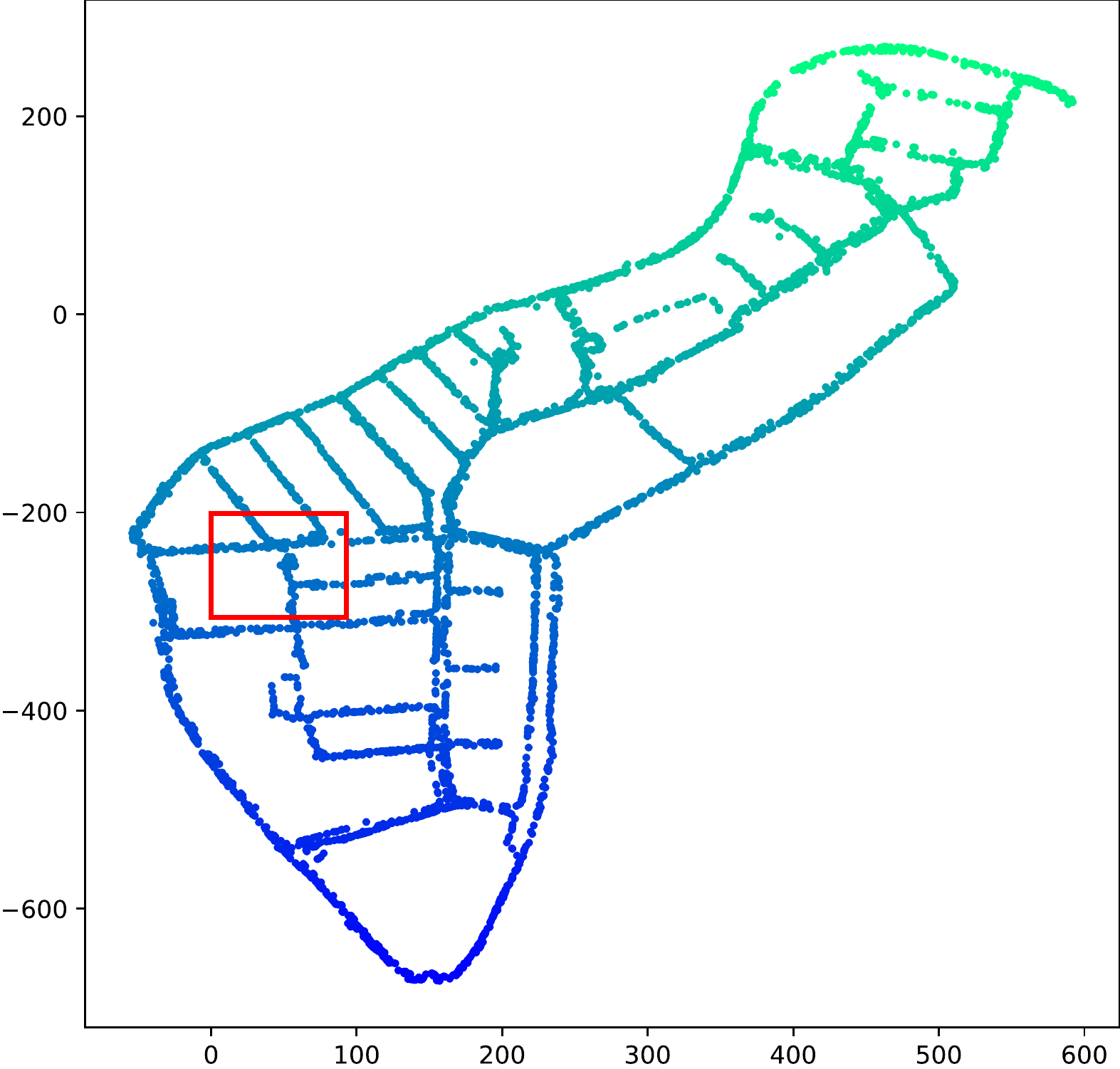}
\par\end{centering}
}\subfloat[NMSE versus the 2-D positions.\label{fig:NMSE}]{\begin{centering}
\includegraphics[scale=0.24]{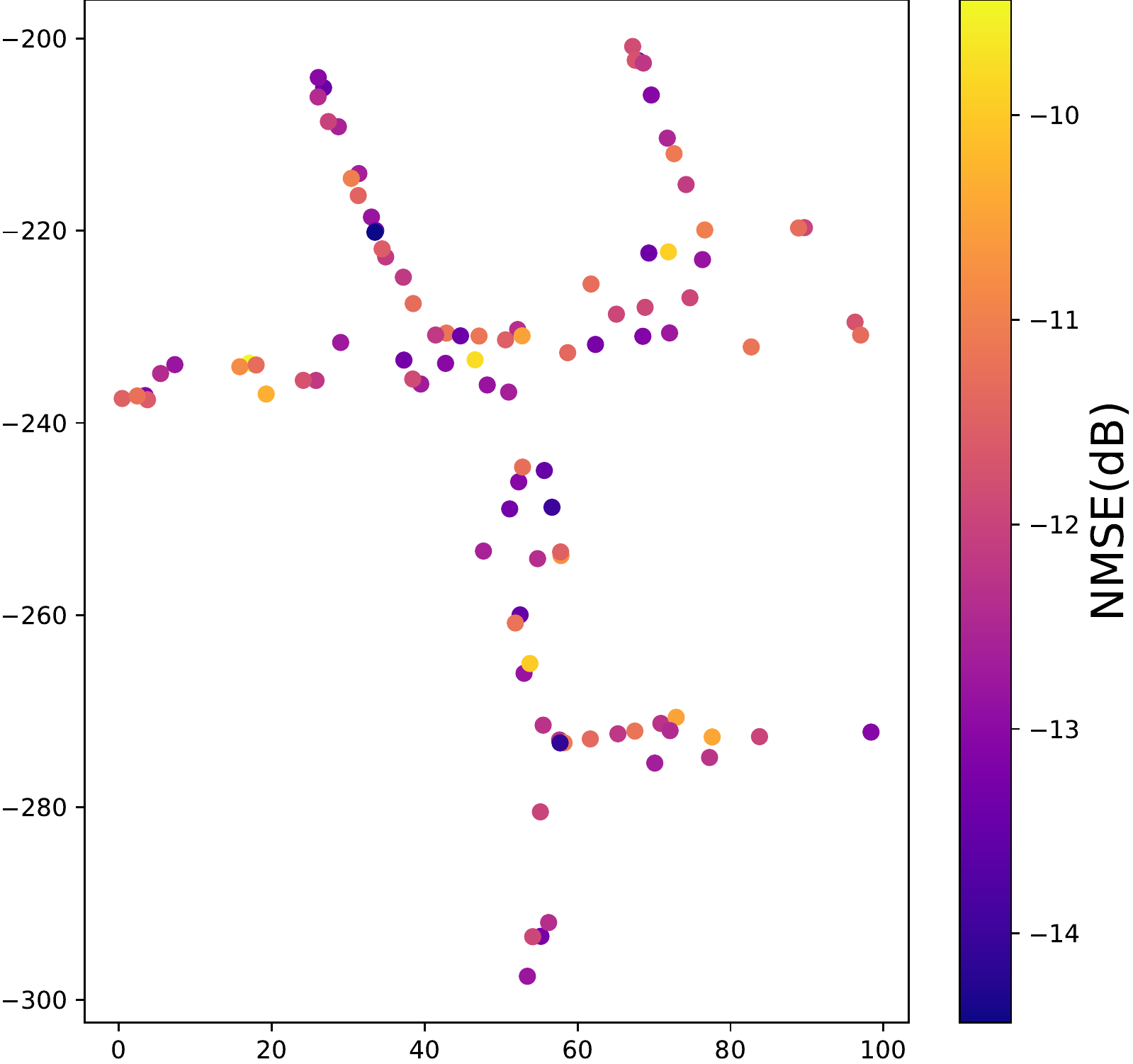}
\par\end{centering}
}

\caption{NMSE distribution over 2D area.}
\end{figure}

Then we consider the NMSE distribution versus the 2D positions. The
dataset adopted is generated from a real residential area, where the
outline of the users' locations is shown in Fig. \ref{fig:Locations}.
We consider a sub-area with size $100\times100$ m$^{2}$, as indicated
by the red box in the Fig. \ref{fig:Locations}, and we randomly pick
100 users to test their CE performance. The NMSE distribution is shown
in Fig. \ref{fig:NMSE}, where we set  $L=24$, SNR = 20 dB and $I_{D}=1500$.
 It shows that the proposed scheme can achieve good CE performance
no matter where the user is located.\textcolor{blue}{}

\section{Conclusions}

In this paper, an algorithm based on CL is proposed for MUCE in massive
MIMO systems. CL is utilized to effectively extract location-embeded
features from the received signal to indicate the correlations among
MU channels. DNet is proposed to exploit the correlations within CSI,
which can be trained jointly with the CLNet to improve the overall
performance. To adapt the proposed scheme to different number of users
and different degrees of correlations, we further propose a subnetwork-based
scheme. Experiments show that our proposed scheme outperforms the
baselines, especially when the pilot length is small and the labeled
data is very limited.


\bibliographystyle{IEEEtran}

\end{document}